\documentclass[10pt,conference]{IEEEtran}
\IEEEoverridecommandlockouts
\usepackage{balance}
\usepackage{graphicx} 
\usepackage[utf8]{inputenc}
\usepackage{amsmath}
\usepackage{amssymb}
\usepackage{amsthm}
\usepackage{physics}
\usepackage{subcaption}
\usepackage{hyperref}       
\usepackage{comment}
\usepackage{fancyhdr}
\usepackage{float}
\usepackage{tikz}
\usepackage{braket}
\usetikzlibrary{quantikz2}

\usepackage[url = false, doi=false, eprint= false, natbib=true, sorting=none, maxcitenames=1]{biblatex}
\usepackage[capitalise]{cleveref}
\usepackage{makecell}
\title{Application-Driven Benchmarking of the Traveling Salesperson Problem: a Quantum Hardware Deep-Dive}
\author{
 \IEEEauthorblockN{Amine Bentellis\IEEEauthorrefmark{1},
 Benedikt Poggel\IEEEauthorrefmark{1}, Jeanette Miriam Lorenz\IEEEauthorrefmark{1}}
  \IEEEauthorblockA{\IEEEauthorrefmark{1}Fraunhofer Institute for Cognitive Systems IKS, Munich, Germany}

 \{amine.bentellis, benedikt.poggel, jeanette.miriam.lorenz\}@iks.fraunhofer.de
}
\date{March 2025}

\renewcommand{\vec}[1]{\boldsymbol{\mathbf{#1}}} 

\fancypagestyle{specialfooter}{%
  \fancyhf{}
  
  \fancyfoot[R]{ \noindent\fbox{%
    \parbox{\textwidth}{%
        {\footnotesize This work has been submitted to the IEEE for possible publication. Copyright may be transferred without notice, after which this version may no longer be accessible.}
        }
    }}
}

\newenvironment{nalign}{
    \begin{equation}
    \begin{aligned}
}{
    \end{aligned}
    \end{equation}
    \ignorespacesafterend
} 
\def\BibTeX{{\rm B\kern-.05em{\sc i\kern-.025em b}\kern-.08em
T\kern-.1667em\lower.7ex\hbox{E}\kern-.125emX}}
\begin{document}

\maketitle
\begin{abstract}

The potential analysis of the capabilities of quantum computing, especially before fault tolerance at scale, is difficult due to the variety of existing hardware technologies with a wide spread of maturity. Not only the result of computations, but also the very process of running quantum-enhanced algorithms differ from provider to provider. The study includes a comparative analysis of various hardware architectures with the example of the Traveling Salesperson Problem, a central class of combinatorial optimization. It highlights what steps are necessary to run real-world applications on quantum hardware, showcases how the providers and various technologies differ and presents results in the relative efficiency of exemplary quantum algorithms on neutral atom-based, ion trap and superconducting hardware, the latter including both gate-based and annealing devices. This is an important step in advancing the understanding of quantum computing capabilities from an application standpoint - agnostic to the underlying qubit technology and projecting results into the future to judge what further developments on the application side are necessary.

\end{abstract}

\thispagestyle{empty} 
\pagestyle{plain}
\thispagestyle{specialfooter} 

\section{Introduction}

Since the proposal of the first theoretical quantum algorithms in the 1990s~\cite{bernstein_quantum_1997, grover_fast_1996, shor_algorithms_1994}, quantum computing (QC) has come a long way. Not only have new algorithmic paradigms that combine classical and quantum computing resources been developed~\cite{cerezo_variational_2021}, but also the practical realization of QC has advanced considerably. An increasing number of hardware providers offer QC access to computers based on different qubit technologies, like superconducting, ion-trap or neutral atom qubits. While this development enables a new practical approach, it also raises the question how the different qubit realizations compare against each other. 

However, an answer is not easily obtained for several reasons: First, the different provider's offers are very diverse in terms of access models, abstraction level, maturity and the ways an end user can control them. Second, they differ not only in size and connectivity, but also in the algorithmic paradigms they support - e.g., thousands of superconducting qubits on a quantum annealer cannot be compared easily with hundreds of qubits on a gate-based superconducting quantum computer. Third, while hardware-specific measures such as coherence times, gate execution times or gate fidelities can be obtained with relative ease, it is entirely unclear how these metrics translate into a benefit for concrete applications. After all, this is the ultimate goal of QC. 

In this study, we provide an application-specific and user-oriented comparison of several types of quantum hardware including superconducting qubits (both annealing and gate-based), ion-trap and neutral atom architectures. This is an important step in asserting which hardware performs at which quality from the point of view of an end user. Furthermore, handling the diversity in access models and the level of control of the QC hardware given by the providers gives a clearer view on what steps one needs to take to solve application problems on real QC hardware.

To this end, we conduct a benchmarking study on the application area of combinatorial optimization. As an exemplary application case, the travelling salesperson problem (TSP) is selected due to its clear problem statement and its relative popularity.  The non-trivial nature of constraints appearing in a QUBO encoding~\cite{palackal_quantum-assisted_2023} places it much closer to a real-world problem than, e.g., a Maximum Cut problem that translates trivially to a QUBO (and therefore Ising) formulation and poses a much smaller challenge because of its unconstrained nature. The solution algorithm is the Variational Quantum Eigensolver (VQE)~\cite{peruzzo_variational_2014} due to its flexibility in the ansatz choice and because it has been extensively studied in the context of noisy hardware. 

After providing a full implementation of the algorithm for the TSP, in some cases modified to accommodate hardware constraints, data on the behavior of the cost function, the scaling of success metrics and the solution quality is collected and finally analyzed. The analysis made as fair as possible even though the disparity of the available data between different hardware is a considerable challenge. Our key contributions are

\begin{itemize}
    \item the collection of experiences, obstacles and ways to overcome them on the road to using real quantum hardware for end users,
    \item showcasing the intricacies of truly application-centric cross-platform benchmarking of quantum hardware and algorithms,
    \item comparing different QC hardware technologies (superconducting, ion-trap, neutral atoms) at their current status of development, with an attempt to extrapolate and generalize the results.
    \item investigating how to solve the TSP on neutral atoms systems and proposing an implementation on real hardware
\end{itemize}

The problem instances used in the study range from 3 to 15 cities where the maximum sizes only fits on superconducting hardware. While the baseline algorithm is VQE with a hardware-efficient ansatz, quantum annealers naturally require a different approach merely encoding the cost function. This is also true for the tested neutral atom system. For testing, both IQM and IBM superconducting hardware was used, as well as a D-Wave quantum annealer, an AQT system based on ion traps, and a neutral atom quantum computer by PASQAL.

To our knowledge, the study and its predecessor on the variational quantum eigensolver~\cite{bentellis_benchmarking_2023} are the first to use the recently increased hardware availability for an estimation of its impact on relevant application cases, taking the whole solution process into account without resorting to an abstract metric like circuit fidelity or quantum volume.

The rest of the paper is structured as follows: \cref{sec:relatedwork} summarizes related work to application-centric benchmarking and the question of advantages and disadvantages of specific hardware realizations. \Cref{sec:background} briefly covers the technical background necessary for the experiments and \cref{sec:digital} and   \cref{sec:analog} presents the result of the comparison. Finally, the implications are discussed together with subsequent research questions in \cref{sec:discussion}.

\section{Related Work}\label{sec:relatedwork}

The relatively recent availability of different types of usable QC hardware has sparked great interest in understanding the prospects of the different technologies for applications. 

There are efforts in establishing an application-centric benchmarking procedure with several tools available~\cite{quetschlich_mqt_2023, finzgar_quark_2022, lubinski_application-oriented_2023}. While they do not tackle the different hardware options explicitly, they are designed to treat the quantum computer from a higher application level, hence enabling indirect comparisons. From a conceptual perspective, possibilities and uses of benchmarking is discussed in~\cite{proctor_benchmarking_2025}. Furthermore, there is an increasing interest in developing a common understanding of QC benchmarking on all levels of abstraction, e.g. from a European perspective~\cite{lorenz_systematic_2025}. Finally, a systematic analysis of the implications of hardware imperfections for variational quantum algorithms is developed in~\cite{barligea_scalability_2025}, raising concerns about the practicability of the paradigm due to the hardness of the classical optimization. However, the paper also establishes a recipe for data-based estimates of variational quantum algorithm performance considering hardware and shot noise. However, we see in all these works a lack of clear hardware comparison study. This work specifically fills that gap by offering a deeper understanding of application-level benchmarking, enabling a better evaluation of the current status of quantum computers.

Looking at applications, countless studies with very specific setups are performed, usually including an attempt at a potential analysis. For different sectors, results are gathered in broader overviews like for finance~\cite{egger_quantum_2020, herman_quantum_2023}, (renewable) energy~\cite{paudel_quantum_2022, giani_quantum_2021} or combinatorial optimization~\cite{abbas_challenges_2024}. The high research interest is an indicator for the urgency with which useful QC applications are in demand, with the technology being increasingly focussed also by economic analysis~\cite{bova_francesco_commercial_2021}. While these attempt a potential analysis for the technology in general, the present study delivers an important ingredient by contributing to a comparison of the potential of different hardware realizations.

Outside of research efforts targeting or making use of a specific hardware (e.g., by IBM research teams), cross-platform benchmarking efforts typically focus on quality metrics close to the hardware layer like fidelities or error rates~\cite{suau_single-qubit_2023, kordzanganeh_benchmarking_2023, zhu_cross-platform_2022, linke_experimental_2017}. An exception is the recent benchmarking of the VQE on different types of hardware~\cite{bentellis_benchmarking_2023}. Even though no conclusive results have been reached, automation efforts already target the hardware selection to ensure the readiness of the software stack or tool pipeline~\cite{weder_automated_2021, phalak_qualiti_2025, quetschlich_mqt_2025}. 

\section{Background}
\label{sec:background}

The necessary ingredients for the study presented in this paper are introduced briefly in this section: the traveling salesperson application, the variational quantum eigensolver, application-centric metrics and the technologies behind the QC hardware used in the experiments.

\subsection{Combinatorial Optimization and the Traveling Salesperson Problem}

Combinatorial optimization is concerned with the selection of a solution from a finite set while minimizing (or maximizing) a given cost function~\cite{du2022introduction}. Typically, the challenge in this problem lies in the sheer amount of possible solutions. A prominent example is the traveling salesperson problem (TSP): Given a set of cities and their pairwise distances, find the shortest cycle that visits each city exactly once~\cite{gutin2006traveling}. Although the optimization problem is NP-hard, approximation algorithms can be successful for many real-world instances, e.~g. for the metric TSP where the pairwise distances satisfy the requirements of a metric (in particular the triangle inequality). In fact, instances with tens of thousands cities are solved to optimality with the Concorde solver~\cite{applegate_traveling_2006}, and approximations go far beyond that. Nonetheless, the simple problem statement makes the TSP a good candidate for the exploration and numerical analysis of optimization algorithms. Furthermore, derived problems such as (capacitated) vehicle routing problem have a high practical relevance and are significantly harder.

In this study, a well-known standard encoding of the TSP as a quadratic unconstrained binary optimization problem (QUBO) is used as in~\cite{palackal_quantum-assisted_2023}. Consider the TSP on a set of $n$ cities labeled $i = 0, \dots, n-1$, and the pairwise distance given in a symmetric adjacency matrix $D$. A candidate solution is then given by assigning the cities to time steps $j = 0, \dots, n-1$ fixing the order in which they are visited. Introducing binary integer variables $x_{ij}$ that indicate whether the statement ``City $i$ is visited at time $j$'' is true, the length of a path is then given by

\begin{equation}
    C_{\mathrm{path}} = \sum_{i, i', j} x_{i,j} x_{i',j+1} D_{ii'}. 
\end{equation}

For ease of notation, periodic time steps are used (i.~e., time $n \equiv 0$). Intuitively, the distance between two cities contributes to the overall path length if the cities are visited at adjacent time steps. Furthermore, it is necessary to introduce constraints to ensure the variable assignment leads to a valid path. At each time step, exactly one city is visited, and each city is visited at exactly one time step. Hence, for each city and each time step, only one of the binary variables associated with it should take the value $1$ with all others being $0$. Encoding these constraints into quadratic penalty terms~\cite{palackal_quantum-assisted_2023} leads to the full cost function

\begin{nalign}
    C(\vec{x}) = &C_{\mathrm{path}} + P \cdot C_{\mathrm{constraints}} \\
    = &\sum_{i,i',j} D_{ii'} x_{ij} x_{i,j+1} \\
    &+ P \sum_i \big(1 - \sum_j x_{ij}\big)^2 + P \sum_j \big(1 -  \sum_i x_{ij}\big)^2.
\end{nalign}

Here, $P$ needs to be chosen large enough to ensure the feasibility of the solution with the minimal energy. Considering that the length of a path does not change with cyclic permutations of the cities, the number of binary variables can be reduced from $n^2$ to $(n-1)^2$ by deriving an effective cost function where city $0$ appears at time step $0$. In this work, this reduced formulation of the TSP is used.

Its formulation stands out by the fact that we need to add constraints to defavour infeasible paths. Only a small fraction of the solution space are feasible TSP paths. Moreover, these constraints are weighted by a penalty term, which in turn  proportionally affect our cost value. That in turn mean that if we consider our optimal cost values, the bigger the penalty the bigger the separation will between our feasible and unfeasible solutions in term of cost. That in turn means that the relevant loss values proportionally stand out in the loss landscape. 

\subsection{Variational Quantum Algorithms}

Variational QC~\cite{cerezo_variational_2021} is a paradigm with the intention of making use of noisy intermediate-scale (NISQ)~\cite{preskill_quantum_2018} quantum hardware by limiting the size and depth of the quantum circuits that need to be executed. This is achieved by constructing a relatively shallow parameterized circuit, the ansatz, and hand it to a classical optimization algorithm with the task to find the optimal parameters under some definition of optimality. 

In the case of the variational quantum eigensolver~\cite{peruzzo_variational_2014}, the cost function of the optimization is given by some Hamiltonian whose expectation value in the state produced by the ansatz is estimated with the aid of a quantum computer. The Hamiltonian may encode a physical system like a molecule, or an optimization problem, e.g. in an Ising model derived from a QUBO expression. 

Formally, the expectation value is given as
\begin{equation}
    C(\vec{\theta}) = \braket{\psi_0 | U^\dagger(\vec{\theta}) H_P U(\vec{\theta}) | \psi_0}.
\end{equation}
Here, $C$ is the cost function, $\ket{\psi_0}$ an easily prepared initial state (often the uniform superposition), $U(\vec{\theta})$ the unitary implemented by the ansatz with parameters $\vec{\theta}$ and $H_P$ the problem Hamiltonian. Once $C$ is minimized, the state $\ket{\psi(\vec{\theta})} = U(\vec{\theta}) \ket{\psi_0}$ encodes a solution to the original problem. In combinatorial optimization, $H_P$ is a diagonal Ising Hamiltonian and the optimal solution is a computational basis state (or the subspace spanned by several equivalent computational basis states).

The success of the variational algorithm crucially depends on the ability of the classical optimizer to find the minimum, or a good approximation for it. Performance differs from case to case, with possible obstacles to convergence being barren plateaus~\cite{mcclean_barren_2018}, local minima~\cite{anschuetz_beyond_2022}, or uncertainties in the cost function estimation~\cite{barligea_scalability_2025}.

\subsection{Comparing Hardware}

The natural metric to assess the success of an algorithm for optimization problems is their cost function $C(\vec{x})$ where $\vec{x}$ is a representation of a candidate solution. However, for real-world application cases, the encoding of the original problem can obfuscate the solution quality from the point of view of the end user, especially when constraints are involved~\cite{palackal_quantum-assisted_2023}. To this end, application-specific performance metrics can be used. For heavily constrained problems such as the TSP in the one-hot encoding, the \emph{feasibility ratio} is useful - it gives the ratio of basis states that satisfy all constraints in the final quantum solution. Clearly, for feasible basis states, the cost function is equal to the TSP path length up to a constant shift and a constant factor. To allow for an easier comparison across different instances of the problem, the difference from the ground state $\Delta_C \coloneqq C(\vec{x}) - C_{\text{min}}$ or the approximation ratio $r_C \coloneqq \frac{\Delta_C}{C_{\text{max}} - C_{\text{min}}}$ are useful. However, this requires that the algorithm is successful in generating feasible solutions (feasibility ratio $\approx 1$). It is important to note that for infeasible states, the QUBO cost function carries no direct meaning when it comes to the actual application.

On a more fundamental level, a typical question concerning the capabilities of current quantum hardware is ``How large are the problems it can solve ? ``. How this question can be answered depends on the technology. E.g., on superconducting gate-based quantum hardware, the number of qubits gives an upper limit to the number of binary variables in the QUBO. However, the connectivity of the problem can require a mapping that results in many SWAP gates, increasing the depth of the circuit. On the other hand, superconducting annealing hardware circumvents the problem of limited connectivity by encoding a single binary variable into many physical ones (the \emph{minor embedding}~\cite{yarkoni_quantum_2022}). This effectively reduces the number of binary variables that can be encoded.

\subsection{Quantum Computing Hardware}

Qubits can be realized in many ways - ultimately, they require a physical two-state system together with the ability to execute operations (including those that involve at least two qubits)~\cite{chae_elementary_2024}. The hardware platforms used in this study are briefly introduced in the following sections. In this work, we focus on both digital (or gate-based) and analog quantum computers. Digital quantum computing implements algorithms through sequences of gate operations whereas analog quantum computing exploits the continuous-time evolution of a controllable Hamiltonian to  emulate a target quantum systems.

\subsubsection{Superconducting qubits}\label{sec:bac_superconducting}
Superconducting qubits rely on the Josephson effect~\cite{josephson_possible_1962} describing the effect of quantum energy levels in a macroscopic system formed by two superconductors separated by a barrier. The most common devices are based on Transmon with systems of considerable size being available, e.g. by IBM~\cite{abughanem_ibm_2025, ibm_authors_ibm_nodate} and IQM~\cite{abdurakhimov_technology_2024}. Drawbacks of the systems are the need for cooling and the fixed arrangement of qubits on a chip which leads to a limited connectivity. 

With Qiskit~\cite{qiskit_contributors_qiskit_2023}, one of the major high-level languages for quantum computing is designed to target superconducting quantum hardware. For this reason, a larger ecosystem has developed into the state where IBM's quantum hardware is seen as the default version of quantum computers. However, this statement can be challenged increasingly for two reasons: First, other providers like IQM are able to deliver systems that come closer in size and quality, and other qubit technologies advance to a point where universal quantum computing is available on multiple platforms.

The superconducting quantum computers built by IQM and IBM implement universal quantum computation in the circuit model. Every unitary transformation between input and output states is realized by applying one- and two-qubit gates to the qubits. Operators involving more than two qubits can be decomposed (compiled) into basis gates. The layout of the hardware furthermore requires a transpilation step that map a theoretical quantum circuit onto the actual machine. This involves both the mapping of theoretical to physical qubits, and the insertion of SWAP gates to realize long-range entangling gates. Ideally, the mapping is done in a way to reduce the swapping overhead and hence the circuit depth.

In this study, the IBM systems \emph{ibm\_fez} and \emph{ibm\_brisbane} were used. \emph{ibm\_fez} is a 156-qubit Heron processor with a 2-qubit error rate of approximately $2\cdot 10^{-3}$ and a median readout error $8\cdot10^{-3}$~\cite{noauthor_ibmresources_nodate}. \emph{ibm\_brisbane} has 127-qubits and processor type Eagle with a slightly higher 2-qubit error rate of $8\cdot 10^{-3}$ and a median readout error of $1.7\cdot 10^{-2}$. Both have a heavy hexagon connectivity with slight differences in the basis gate set.

As an alternative, the study benchmarked the 20-qubit \emph{iqm\_garnet} device~\cite{noauthor_iqmgarnet_nodate}. It follows a simple square-lattice connectivity and reaches median 2-qubit error rates around $5\cdot 10^{-3}$.

\subsubsection{Quantum annealing}
\label{sec:annealing}

D-Wave's quantum annealer~\cite{mcgeoch_advantage_2022} are based on superconducting transmon qubits as well, but do not implement universal quantum computing, but instead quantum annealing, a specialized algorithm to solve optimization problems. 
By gradually changing the system parameters, the Hamiltonian can transition from a Hamiltonian $\mathcal{H}_t$ with a groundstate that is easily prepared (e.g. the uniform superposition of all states),  to $\mathcal{H}_p$, the problem Hamiltonian. The evolution needs to avoid rapid changes that would cause the system to leave its ground state. 

Quantum annealing processors inherently yield low-energy solutions, aligning perfectly with the goals of combinatorial optimization applications that seek to identify configurations with minimal energy. The idea and implementation exist for a long time now, and has seen use in multiple domains \cite{Yarkoni2022, Neukart2017}. Moreover, looking in the direction of the TSP, it seems to have been exactly solved up to 8 cities \cite{10.3389/fphy.2021.760783}. 

D-Wave offers a quantum annealer with approximately $5000$ qubits and $40000$ couplers~\cite{mcgeoch_advantage_2022} (for comparison, full connectivity would require about $12.5$ million couplers). The limited connectivity is handled by encoding the binary variables of an optimization not on individual qubits, but groups of them (\emph{minor embedding}). On top of the quantum system, a selection of hybrid quantum-classical solution methods relying on a decomposition approach is offered~\cite{d-wave_developers_d-wave_2020}. 

\subsubsection{Ion Traps}

Another realization of qubits tested within this study relies on qubit states encoded in individual ions. They can be controlled using electromagnetic fields with the possibility to shift them in in space to realize an all-to-all connectivity. Machines are developed, e.g., by IonQ~\cite{chen_benchmarking_2024} and AQT~\cite{carina_ion-trap_2024}. For this study, a 20-qubit system by AQT is used featuring all-to-all connectivity and 2-qubit error rates around $6\cdot 10^{-3}$~\cite{carina_19-inch_nodate} is used. Although ion traps generally face more challenges in scaling with respect to the number of qubits, they offer higher gate fidelities compared to other technologies, as well as longer coherence times, making them less susceptible to noise. Moreover,  quantum algorithms can be formulated as gate-based circuits which facilitates the high-level access. The quantum processors from AQT are also operated using qiskit.

\subsubsection{Neutral atoms}

Furthermore, qubits can be encoded into the electronic spin states of neutral atoms. The atoms are trapped and manipulated with laser and/or microwave pulses, with good stability and high coherence times being a particular advantage of the technology~\cite{wintersperger_neutral_2023}. For this study, a PASQAL system was tested, though both PASQAL~\cite{noauthor_pasqal_nodate} and QuERA~\cite{noauthor_quera_nodate} have started to provide systems with tens to hundreds of qubits. However, their abstraction level towards end users is not as advanced as for superconducting qubits, and running experiments requires more manual tuning.

Neutral atoms arrays are suitable for implementing quantum Hamiltonians using their Rydberg atoms array. For example, the Ising model can be formulated as \cite{Henriet2020quantumcomputing}: 

\begin{equation}\label{eq:pasqalham}
\mathcal{H}(t)=\frac{\hbar}{2} \Omega(t) \sum_j \sigma_j^x-\hbar \delta(t) \sum_j n_j+\sum_{i \neq j} \frac{C_6}{r_{i j}^6} n_i n_j
\end{equation}

with $n_j = (1 + \sigma^{z}_{j})/2$ the state occupancy and $\sigma^{x,z}_{j}$  the Pauli matrices $\sigma^{x,z}$ of the spin $j$. : this \textit{drive} Hamiltonian governs a time-dependent evolution that depends on the pulse shape, i.e. the Rabi frequency $\Omega(t)$  and detuning $\delta(t)$. \cref{fig: pasqalseq} presents an example of a pulse shape in which the two parameters are adjusted while ensuring they remain within their physical value limits.
Moreover, when constructing algorithms to solve problems on a neutral atom system we also have control on the placement of the atoms on a lattice. The positioning has a big influence since the coupling for the last term in \cref{eq:pasqalham} strongly depends on the distance $r_{i j}$ between qubits i and j.

Solving combinatorial optimization problems using neutral atoms appears to be a promising approach. Notably, important problems such as the Maximum-Independent Set (MIS) are particularly well-suited for this hardware \cite{PhysRevResearch.6.023063, doi:10.1126/science.abo6587,        wurtz2024industryapplicationsneutralatomquantum}. This stems from the simplicity of directly embedding the MIS graph onto the QPU, as the independent set constraint is inherently satisfied by the Rydberg blockade mechanism
\cite{Henriet2020quantumcomputing}. Many approaches proposed using the QPU in a sampling procedure to get good solutions to the MIS and using it as subroutine to larger algorithms.
When it comes to solving the problems on the QPU, however, the main candidates are variational algorihtms, notably, QAOA \cite{bauer2024solvingpowergridoptimization, doi:10.1126/science.abo6587} and varitional quantum adiabatic algorithm (VQAA) \cite{coelho2022efficientprotocolsolvingcombinatorial}.
Nonetheless, to the best of our knowledge, there has been no endeavor to address the TSP using neutral atom quantum systems. This absence of attempts may arise from the inherent difficulties associated with solving constrained optimization problems on such hardware \cite{10.3389/fphy.2014.00005}, or possibly from the intrinsic incompatibility of the TSP with neutral atom-based computation methods. 

\section{Digital quantum computers}\label{sec:digital}

In this section, we analyze and compare the results obtained from various quantum computers across different providers, all operating within the same digital quantum computing framework.
For gate-based paradigms, the comparison can be rather straight forward. One needs to make sure that the circuit and setup are similar. Concretely, the study aims for a fair comparison with the base parameters given in the following section.

\subsection{Experimental setup}\label{sec: expsetup}

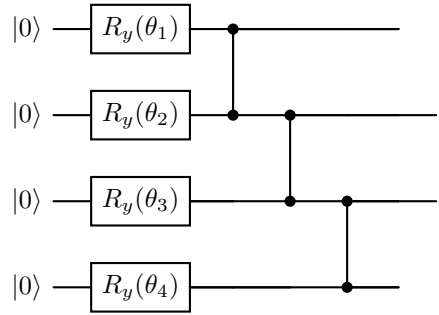
\begin{figure}[htpb]
\centering
\begin{quantikz} 
\centering
\lstick{$\ket{0}$} & \gate{R_y(\theta_1)} &  \ctrl{1} &\qw&\qw&\qw \\
\lstick{$\ket{0}$} & \gate{R_y(\theta_2)}& \phase{}  & \ctrl{1}&\qw&\qw&\\
\lstick{$\ket{0}$} & \gate{R_y(\theta_3)}&\qw& \phase{}& \ctrl{1}&\qw &\\
\lstick{$\ket{0}$} & \gate{R_y(\theta_4)}&\qw&\qw& \phase{} &\qw
\end{quantikz}
\caption{\label{fig:fig_hwery} 4 qubits example of a RY-CZ ansatz}
\end{figure}

The experimental settings chosen for all the runs in this study are:
\begin{itemize}
    \item Circuit: RY-CZ (c.f. \cref{fig:fig_hwery})
    \item Optimizer: NFT \cite{PhysRevResearch.2.043158}
    \item Number of shots: 200
    \item No error mitigation
\end{itemize}    

To ensure a fair comparison among all the hardware types presented in this study, we choose not to implement any error mitigation techniques (or any other error suppressing methods), allowing us to observe the impact of hardware noise on our results. 

The circuit was selected based on empirical testing with a simulator to identify the most effective ansatz for this specific problem. After transpilation, the only variation will be in the gate counts, which depend on the native gateset of the hardware. Indeed, the circuit is simply transpiled into the different basis gates defined by sets dependent on the hardware. IQM Garnet has X and Y rotation as well as CZ for the entangling gate, whereas IBM Heron uses X and Z rotation gates. On the other hand AQT Marmot uses X and Z rotations but RXX as the non-local gate. That makes the transpiled circuit slightly different for the three backends regarding gate count, but the depth stays the same in all cases. The choice of optimizer is influenced by previous testing \cite{bentellis_benchmarking_2023,} and also for its noise resilience \cite{barligea_scalability_2025}.

The decision regarding the number of shots was influenced by the fact that the AQT Marmot system accessed via the LRZ is limited to 200 shots at the moment of the study (which is not the case for AQT's cloud access). For the purposes of this study, and given the small number of qubits, this shot count was adequate for distinguishing hardware noise across different systems. An alternative approach could have been to execute the same circuit multiple times to go past this shot limit; however, this would result in varying levels of inherent drift noise across different runs and providers, making it difficult to account for. Furthermore, at small problem sizes, good results need to be achieved with a relatively low number of shots.

\subsection{Results}

 We are considering the variations in performance between different superconducting qubit hardwares (IQM, IBM) and ion trap system (AQT). This is primarily done through the lens of resulting VQE energies and metrics such as the feasibility ratio, to give an account of the hardware noise's effect on both the optimization process and on its solution.  

\begin{figure}[htbp]
    \centering
    \includegraphics[width=0.45\textwidth]{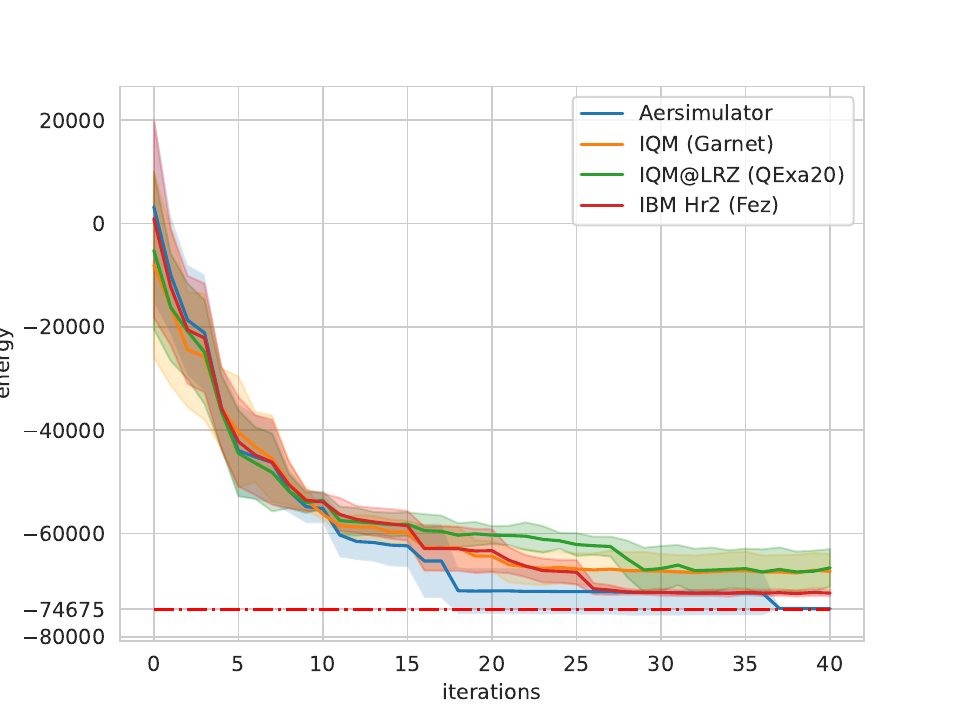}
    \caption{Optimization process comparison between IBM and IQM chips on a 5 nodes TSP (16 qubits). For all experiments the setup is identical: same parameter seeds, optimizer, number of shots, time between experiments}
    \label{fig: iqmcomp}
\end{figure}

We access the most up to date hardware that would be able to solve our problem, from IQM and IBM. Since AQT's hardware was limited to 12 qubits, it was not possible to conduct any experiments involving the 5-node TSP on the ion trap hardware.
 \cref{fig: iqmcomp} compares the energy in the optimization process until convergence for both provider (and using qiskit AerSimulator as baseline)  on three parameter seeded VQE runs over the same graph. Two types of access were available for IQM, via cloud and via the HPC center at the Leibniz-Rechenzentrum; both methods of access are presented here for comparison. Accessing quantum hardware through HPC significantly speeds up variational algorithms, as the integration between classical and quantum systems is more closely established in these environments than it is with cloud access. The simulator results demonstrate that, with this specific setup, parameter seeds and  shot count, the NFT optimizer successfully converges to the exact ground state energy (indicated by the red dotted line).
In \cref{fig: iqmcomp} we observe that the IBM hardware has an advantage in terms of final energy, but more notably, it exhibits less uncertainty, as the results from the Heron r2 chip consistently remain close to the optimized energy. This observation cannot be solely explained by the reported gate error rates for both providers at the time of this study (\cref{sec:bac_superconducting}), highlighting the difficulty of inferring result quality directly from error metrics. This can neither be explained by differences in the circuits after transpilation, as the circuit used (\cref{fig:fig_hwery}) contains fewer gates on the IQM hardware.

\begin{figure}[htbp]
    \centering
    \includegraphics[width=0.45\textwidth]{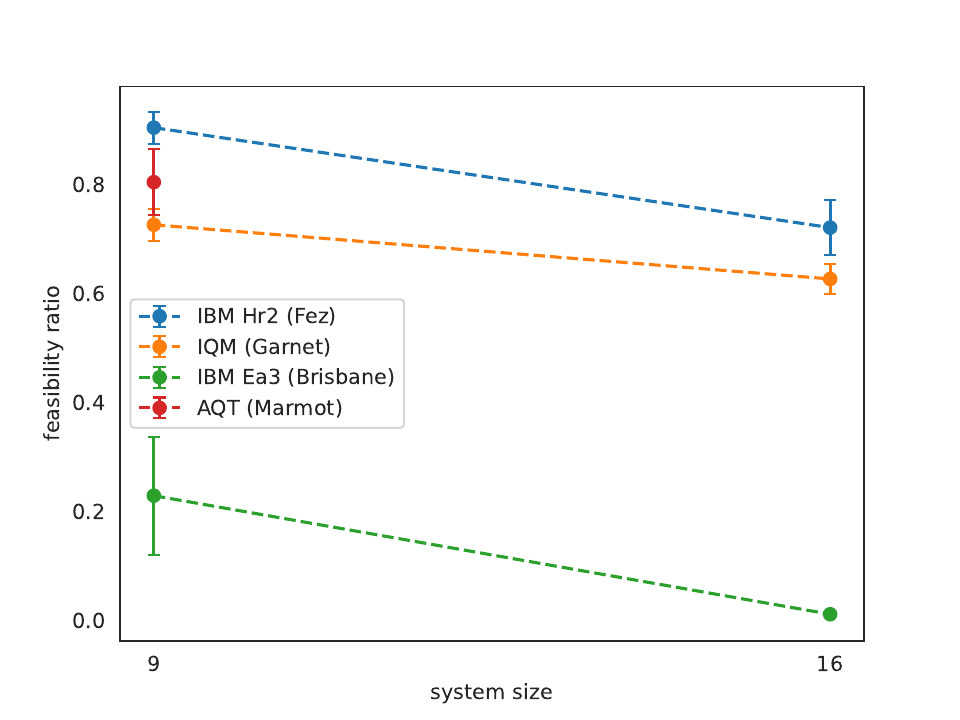}
    \caption{Feasibility ratio comparison between multiple providers over three runs. This experiment was performed using 200 shots as explained in \cref{sec: expsetup}.}
    \label{fig: feascomp}
\end{figure}

The application-specific performance metric shown in \cref{fig: feascomp} illustrates the comparison of feasibility ratios on the same optimal parameter point obtained from noiseless VQE simulations. For example, across 200 sampled solutions by the IBM Heron r2 chip $\sim 90\%$ were feasible solutions. This provides a reliable indication of hardware noise, as the value should be 1.0 if noise was not present (even for shot-based noiseless simulations).
The ratio of feasible states appears to correlate with the lower energy values attained by the IBM Heron r2 hardware.  As expected, as we increase the problem size, here from 4 to 5 nodes TSP, we see a drop in the number of feasible states sampled for all hardwares. This can be attributed to the fact that the solution space is larger, providing more opportunities for hardware noise to compromise the results through random bit flips. Most values lie around 0.75 to 0.85, however, the older IBM chip, \textit{ibm\_brisbane}, performs significantly worse. Even though both chips are currently available to be accessed on the IBM cloud, the results from the Eagle r3 chip are subpar, delivering only a small fraction of feasible paths. Again, this difference cannot only be accounted to the error rates as they lie in the same order of magnitude (see \cref{sec:bac_superconducting}).Another possibility could be the difference between the two circuits after transpilation, as IBM Heron r2 contains fewer gates due to CZ being native to the system. However, as observed in the previous results with IQM, it would be difficult to attribute the difference in outcomes solely to this factor. This can stem from the \textit{signal-delivery innovations} advertised by IBM, even though the inner workings of this innovation are not clearly outlined.

\cref{tab:res_simulation} give a general overview and comparison of the best attained energies by different hardware and their respective feasibility ratio. The Heron r2 chip seem to outperform the other chips in all application-relevant metrics. Again we see a clear gap between the two IBM chips, showing a technological advancement from generation to generation. 

\begin{table*}[htpb]
    \caption{Quantitative comparison of the tested quantum processors on a 9 qubits or 4 node TSP problem}
    \centering
    \begin{tabular}{|l | l | l |}
    \hline    
          & Best energy out of 3 runs   & Feasibility ratio (\%)   \\ \hline
        \textit{ibm\_fez}  & $-14451.9675 $    & $90.5 $ \\ \hline
        \textit{ibm\_brisbane}  & $-10002.7341$ & $22.9$ \\ \hline
        \textit{iqm\_garnet}  & $-13443.8411$ & $72.6 $ \\ \hline
        \textit{aqt\_marmot} & $-13770.8199 $ & $80.5 $ \\ 
        \hline
    \end{tabular}
    \label{tab:res_simulation}
\end{table*} 

These results shows a clear difference between hardware capabilities on the TSP using VQE and shallow circuits. The choice of circuit, optimizer and hyperparameters were done as to not give an advantage to any connectivity or hardware type. These results would be different if, for example, the circuit would include long range entanglement which would give a clear advantage to AQT Marmot's all-to-all connectivity or IQM's square lattice. This is an important distinction because, inversely, denser connectivity also increases the channels where cross-talk between qubits can happen.
\section{Analog quantum computers}\label{sec:analog}

In contrast to digital quantum computing, which is based on discrete qubits and gate operations, analog quantum computing functions on continuous quantum states and utilizes continuous transformations. Thus, the algorithms used differ a lot from the ones used in the previous section which makes the \textit{exact} comparison of digital and analog algorithms an ill-defined task. 

\subsection{Neutral atoms}

Neutral atom devices are unique in how the algorithm can be implemented and encoded on the hardware. They operate in two computational modes. The first is a gate-based mode, where sequences of laser pulses are directed at specific qubits to create the gates.  This is reminiscent of the previous sections and other types of hardware. Individual qubits can be addressed by tightly focused laser beams - optical tweezers. Given that the distance between atoms in the register is usually several micrometers, it is possible to target specific qubits with a high degree of precision. However, current neutral atom providers like QuERA and PASQAL favorize operating in analog mode. This mode is characterized by the simultaneous addressing of the entire qubit array while gradually evolving the system toward the ground state of the target Hamiltonian through the design of appropriate pulse shapes (see \cref{sec:background}).

The methods we choose to solve the TSP, considering the literature, are the variational quantum adiabatic algorithm (VQAA) and QAOA algorithms. Unfortunately, early results showed a clear unsuitability of QAOA for this problem, as the algorithm could not converge to any meaningful solutions. Thus, the VQAA will be explored here more thoroughly. This approach, similar the variational quantum algorithms (VQA), involves a classical optimizer, more specifically Bayesian optimization (BO), to determine the pulse shape, and employs an iterative process to refine the pulse sequence and minimize our cost function.

\begin{figure}[ht]
    \centering
    \includegraphics[width=0.3\textwidth]{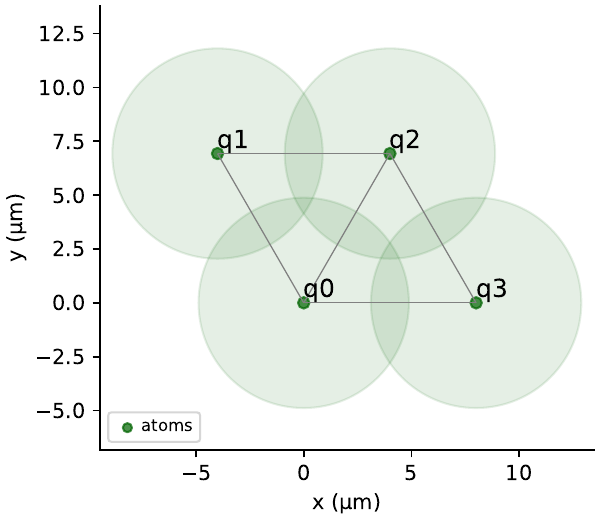}
    \caption{Example of an atom placement for a 3 node TSP.The disk surrounding each qubit signifies the half-radius and indicates the region where coupling between two atoms is feasible.}
    \label{fig: pasqalatom}
\end{figure}

When investigating the QUBO formulation in the context of neutral atom array placement, we discovered that there is no universal encoding for general problems or heuristic best placement; each graph instance requires a different qubit arrangement. This poses a challenge, as this involves solving an NP-complete problem, which consequently introduces a significant classical computational overhead. The main approach involves encoding the off-diagonal components of the QUBO matrix through the Rydberg interaction between atoms (\cref{eq:pasqalham}). \cref{fig: pasqalatom} gives an example placement for a problem requiring 4 qubits.

\begin{figure*}[h]
    \centering
    \includegraphics[width=\textwidth]{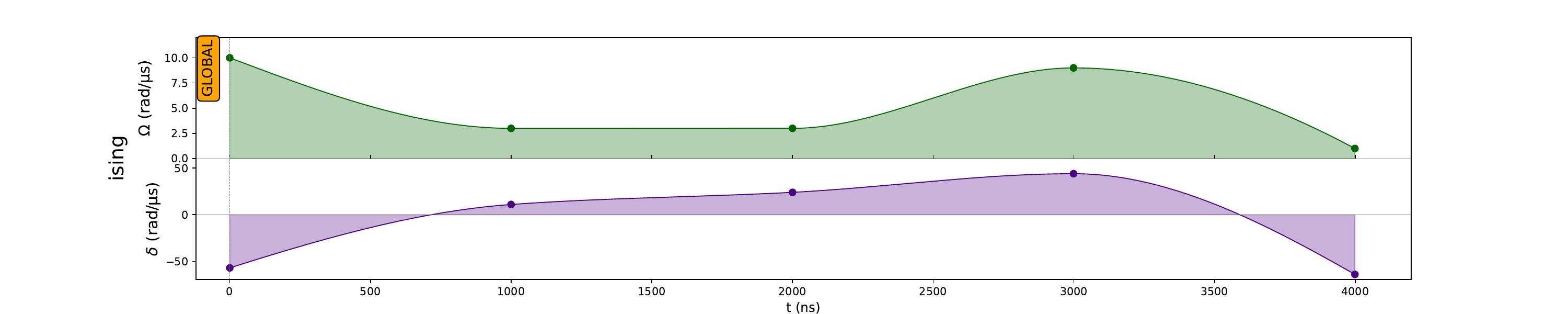}
    \caption{Example pulse sequence to solve one TSP instance. This pulse sequence is optimized using BO, however, it only fits graph instance, so we cannot generalize it to other same size instances. The evolution is done over 4000 nanoseconds, the maximum allowed by PASQAL's chip}
    \label{fig: pasqalseq} 
\end{figure*}

Since we have no prior knowledge of what the pulse shape should be, we optimize it in a similar way as the variational circuits. We take 10 points across the whole evolution (5 for each $\Omega$ and $\delta$), and we interpolate between these points while a \textit{black-box} BO procedure finds the best point configuration by minimizing our cost function (\cref{fig: pasqalseq}). The underlying BO pipeline is similar as in \cite{PhysRevResearch.6.023063}.

\begin{figure}[h!]
    \centering
    \includegraphics[width=0.49\textwidth]{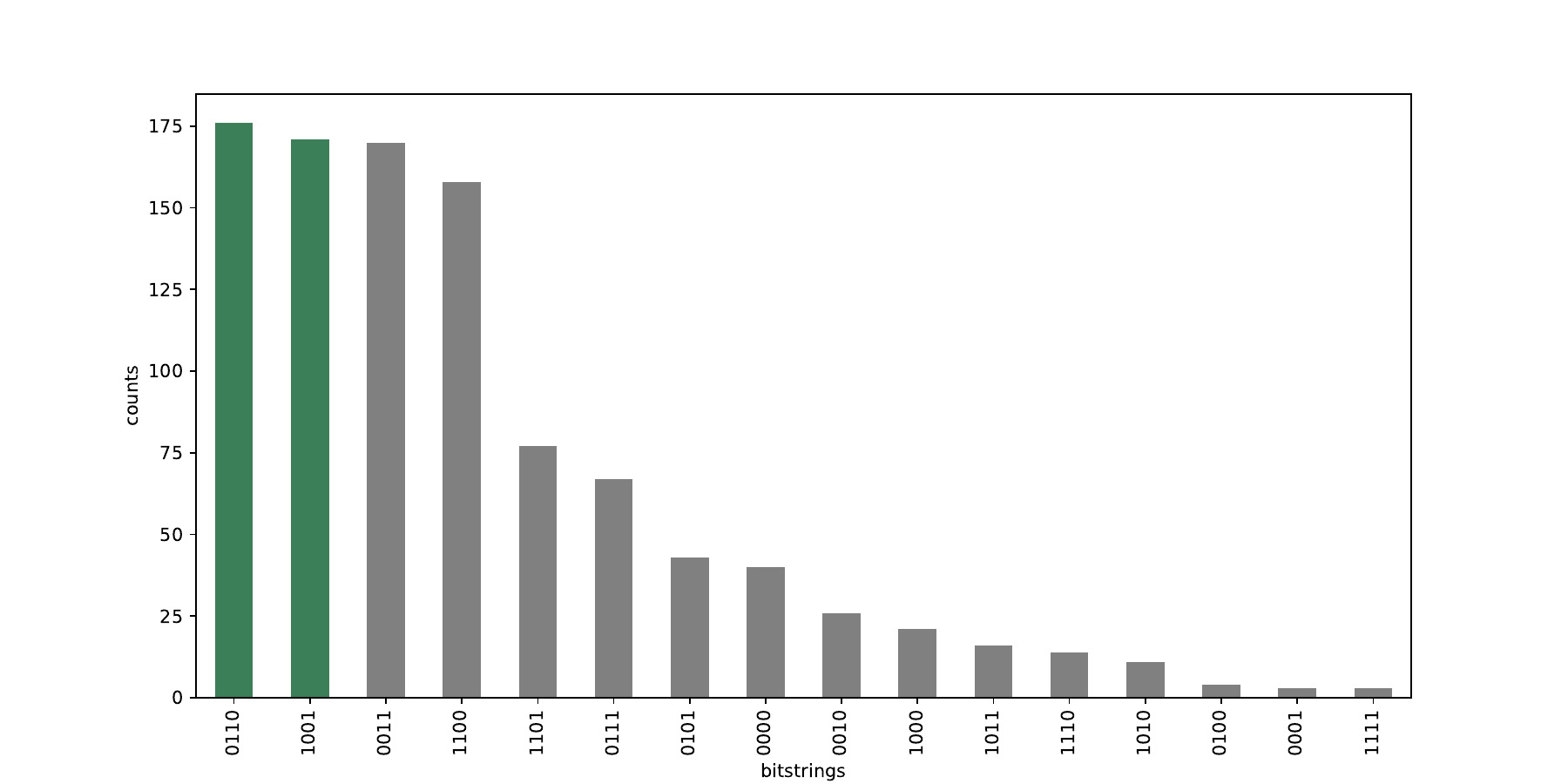}
    \caption{Results for the 4 qubit instance of TSP. In color green 
 we see the optimal solutions (same path but different starting point) and in grey are the unfeasible solutions.}
    \label{fig: pasqalres}
\end{figure}

In \cref{fig: pasqalres}, we show results for a 3-node graph. The algorithm's performance declines significantly as the number of nodes in the graph increases—beginning at 9 qubits, it starts to struggle with finding feasible solutions. There might be multiple reasons, beyond the hardware noise levels, for the subpar results compared to other hardware types that manage to solve bigger TSP instances than 3 nodes. First, the absence of a digital mode, which prevents individual qubit addressing, significantly limits our selection of algorithms. Secondly, the PASQAL QPU is locked, at the moment of our the study, to a triangular lattice shape, which means that we cannot freely optimize the placement of our atoms. This is especially crucial for constrained problem like TSP where we might want to encode the constraints directly in the placement of the atoms. Third, a 4000 nanoseconds cap for the time evolution on the hardware which makes the results not as good as the simulator. Similar to most adiabatic processes, a slower evolution typically yields better results.
Moreover, it also limits how much we can adjust the pulse shape due to the physical constraints on how the laser can vary.

\subsection{Quantum annealing}

\begin{figure}[!h]
    \centering
    \includegraphics[width=0.45\textwidth]{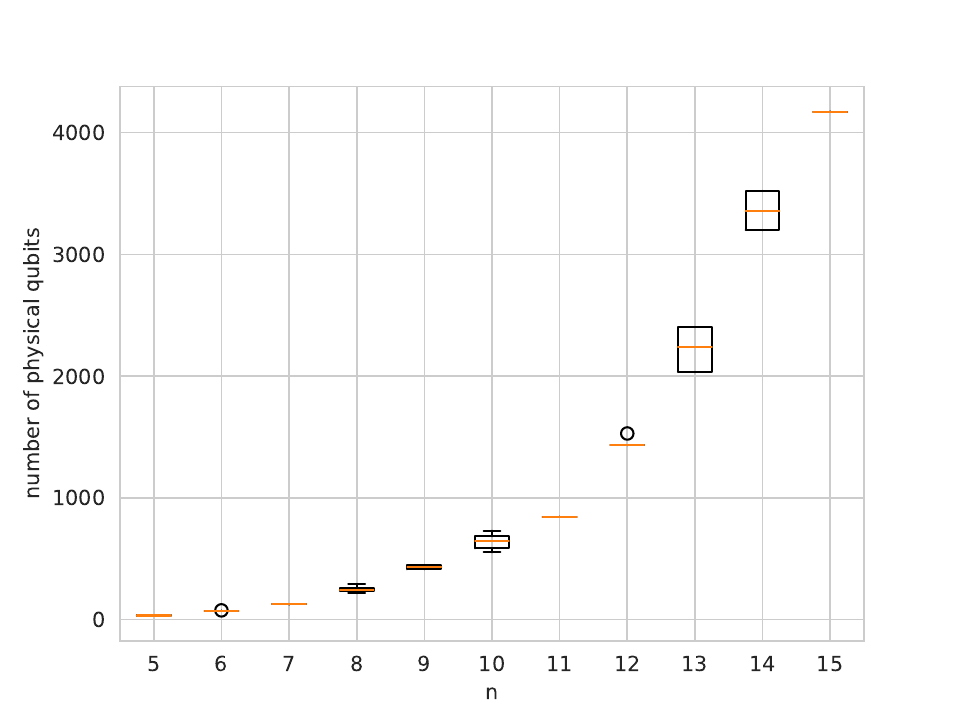}
    \caption{Scaling of the number of physical qubits used against the number of graph nodes.}
    \label{fig: dwavequbits}
\end{figure}

Here we report results from solving the TSP using D-Wave's proprietary methods and systems. For the anneal scheduling we used the default setting that the hardware uses. 
Another parameter we need to consider is the encoding of the problem to the physical qubits. Indeed, the QPUs of D-Wave's Advantage2\textsuperscript{\texttrademark} do not possess all-to-all connectivity, which makes the embedding crucial for good performance. The main challenge with implementing Hamiltonians on a complete graph lies in the inefficiency of embedding complete graphs onto a chimera graph \cite{Choi2010}. Our formulation of the TSP already scales quadratically with the number of nodes in the graph. On top of that, the minor embedding, i.e., the encoding from variable to physical qubit, adds extra overhead to the total number of qubits. \cref{fig: dwavequbits} illustrates this point by depicting the number of physical qubits as function of the graph size for up to 15 nodes. The largest D-Wave backend possesses 5627 qubits and 40279 couplers. From these figures we can conclude that the current embedding schemes will not work with size above 15 graph nodes. This calls for either decomposition methods or a different encoding of the TSP that allows for a optimized minor embedding process.  

\begin{figure}[!h]
    \centering
    \includegraphics[width=0.45\textwidth]{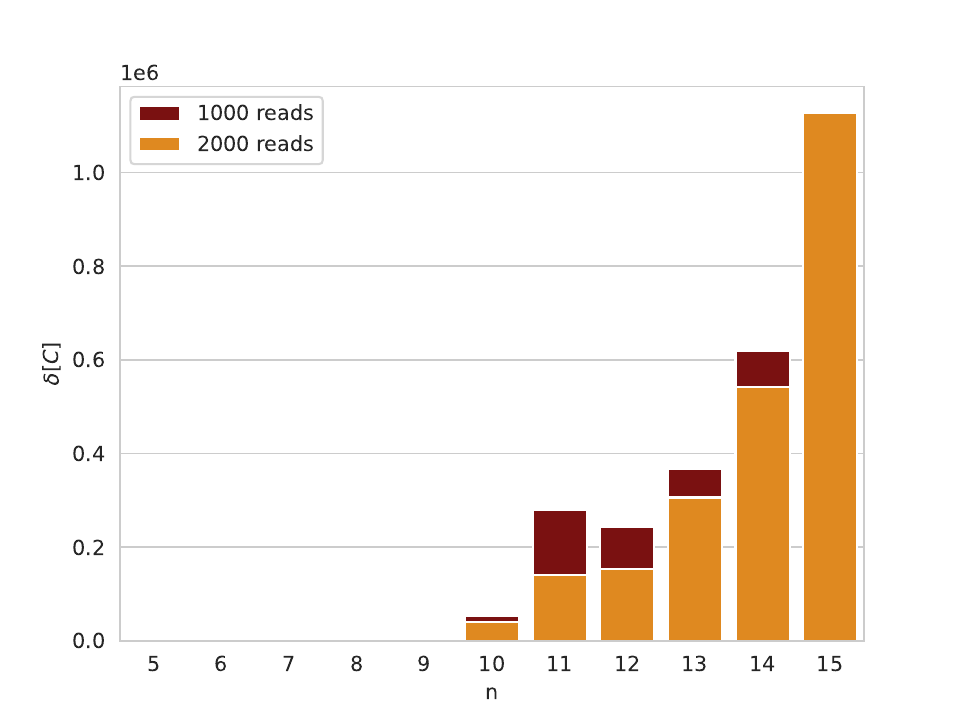}
    \caption{Absolute cost difference (note the scaling on the top left) between the exact solution computed classically and the quantum solution. The number of reads are expressed in different colors and stacked onto each other.}
    \label{fig: dwaveres}
\end{figure}

Due to the maturity of this hardware compared to the previous ones, we can solve larger problems. \cref{fig: dwaveres} presents the difference between the best classically computed solution and the best solution found by the quantum processing unit (QPU) with different numbers of shots. Since this method is still based on sampling the quantum state to get the solution, running it multiple times increases the likelihood of hitting a solution closer to the optimal one. From graph sizes 5 to 9, the optimal solution is always reached. However, as we increase the number of nodes in the graph, it becomes exponentially harder to find the optimal solution.

\section{Discussion}\label{sec:discussion}

From our research, we can draw several conclusions. First, the results from IBM hardware and the advancements seen across chip generations indicate that hardware development is progressing positively. As seen in these results, progress is not limited to the reduction of error metrics alone but also includes a broad spectrum of techniques that can be aimed at mitigating or suppressing errors. Moreover, if we follow the roadmaps outlined by leading companies in hardware development, we will soon possess the capability to tackle large-scale problems.  Secondly,  while the maturity of each hardware platform varies, each provider has distinctive features that could be leveraged using the right algorithm. This was showcased throughout this work, for example, neutral atom's capability to encode the problem onto the connectivity of the qubits or integrating a quantum computer into an HPC environment. These features can enable focused, and more importantly, \textit{useful} algorithm design tailored around them. Third, there is a need to explore a wider array of problems, ranging from the more mathematical ones to the more practical like the TSP. The existing literature, particularly regarding neutral atoms, primarily addresses unconstrained and hardware-efficient problems like the MIS. While the synergy between hardware and algorithm design is crucial, it is also important to define which problems are implementable or not for specific hardware. Identifying the failure modes of hardware in relation to particular applications is another step towards achieving practical quantum computing. Fourth, from an application point of view, efforts that attempt to unify the access to different types of hardware are of extreme important. With multiple providers and hardware technologies reaching chip sizes that allow for benchmarking, the amount of hardware- and vendor-specific adaptation needs to be reduced to realize efficient processes. In the same way, reaching a common definition of the figures of merits and the goals to achieve would help producing more direct comparisons. 

\section{Conclusion}\label{sec:conclusion}
\balance
As the field of quantum computing moves toward solving meaningful real-world problems, there is an increasing effort to understand the contexts in which quantum computing, and by extension quantum hardware, can provide added value. It may seem obvious, but  \textit{effectively applying} quantum computing necessitates a solid understanding of quantum hardware, not just the algorithms themselves. Thus, the co-design of hardware and algorithms appears to be essential. This work builds upon previous studies focused on addressing meaningful challenges and assessing the readiness of current hardware.

In conclusion, we are still primarily focused on toy-sized problems, not only because of hardware limitations but also due to the inadequacy of current algorithms in terms of specialized hardware compatibility. For example, variational algorithms aim to provide early advantages for all types of hardware, yet their dependence on classical optimization methods amidst different noise sources poses significant challenges, as optimizers struggle to perform effectively under such conditions \cite{barligea_scalability_2025}.

This benchmarking effort also aims at clarifying the different views amongst the community around the stages of algorithm development.
Advocates for NISQ algorithms emphasize the importance of gaining early advantages, while others argue that the hardware is not yet mature enough, and that theoretical algorithm development should be the community's sole focus. A more insightful perspective would be to recognize that the two modes exist on a continuum shaped by advancements in hardware. A pertinent question for NISQ proponents is what the implications of their research will be once fault-tolerant qubits become available. Conversely, for those focused solely on fault-tolerant algorithms, how adaptable will their approaches be in a future where both fault-tolerant and noisy qubits coexist? This discussion underscores the necessity for \textit{continuous} benchmarking the whole stack (hardware to application) to assess hardware readiness for relevant problems and allows for comparisons to determine which hardware may be more effective for specific applications.

\section*{Acknowledgment}
We thank the Leibniz Supercomputing Centre (LRZ) to have provided access
to the Q-Exa and AQT systems, and support during the test phases of
these machines.
This research is part of the Munich Quantum Valley, which is supported by the Bavarian state government with funds from the Hightech Agenda Bayern Plus.
\clearpage
\balance
\printbibliography
\end{document}